\documentclass[twocolumn,twoside]{IEEEtran} 

\usepackage{amsmath}
\usepackage{amssymb}
\usepackage{array}
\usepackage{fixmath}
\usepackage{graphicx}
\usepackage{cite}
\usepackage{color}
\usepackage{changes}
\usepackage{algorithm}
\usepackage{algorithmic}
\usepackage{blindtext}
\usepackage{acronym}
\usepackage{soul}

\DeclareMathAlphabet{\mathbit}{OML}{cmr}{bx}{it}

\newcommand{\B}[1]{{\bf{#1}}}
\DeclareMathOperator{\Transpose}{T}
\DeclareMathOperator{\Hermitian}{H}
\newcommand{\Tr}{{\Transpose}}
\newcommand{\He}{{\Hermitian}}

\DeclareMathOperator{\atan}{atan}

\DeclareMathOperator*{\argmax}{argmax}

\newcommand{\eul}{{\text{e}}}
\newcommand{\imj}{{\text{j}}}

\usepackage[numbers,sort&compress]{natbib}

\acrodef{ADC}[ADC]{Analog-to-Digital Converter}
\acrodef{AO}[AO]{Alternating Optimization}
\acrodef{AoA}[AoA]{Angle of Arrival}
\acrodef{AoD}[AoD]{Angle of Departure}
\acrodef{APN}[APN]{Analog Precoding Network}
\acrodef{ASK}[ASK]{Amplitude-Shift Keying}
\acrodef{AWGN}[AWGN]{Additive White Gaussian Noise}
\acrodef{BER}[BER]{Bit Error Ratio}
\acrodef{BF}[BF]{Beamforming}
\acrodef{BFN}[BFN]{Beamforming Network}
\acrodef{BS}[BS]{Base Station}
\acrodef{CB}[CB]{conjugate beamforming}
\acrodef{COMP}[COMP]{Covariance OMP}
\newacro{CSI}{Channel State Information}
\acrodef{DAC}[DAC]{Digital to Analog Converter}
\acrodef{DBS}[DBS]{Distance-Based Scheduling}
\acrodef{DCS}[DCS]{Digital Communication System}
\acrodef{DCOMP}[DCOMP]{Dynamic COMP}
\newacro{DFT}{Discrete Fourier Transform}
\acrodef{DL}[DL]{downlink}
\acrodef{DOA}[DOA]{Direction Of Arrival}
\acrodef{DPC}[DPC]{dirty-paper coding}
\acrodef{ESD}[ESD]{Energy Spectral Density}
\newacro{FDD}{Frequency-Division Duplex}
\acrodef{FSK}[FSK]{Frequency-Shift Keying}
\acrodef{FT}[FT]{Fourier Transform}
\acrodef{HT}[HT]{Hilbert Transform}
\acrodef{HI}[HI]{Harmonic Interference}
\acrodef{ICI}[ICI]{Inter-Carrier Interference}
\acrodef{IL}[IL]{Insertion Losses}
\acrodef{ISI}[ISI]{Inter-Symbol Interference}
\acrodef{JSDM}[JSDM]{Joint Spatial Division and Multiplexing}
\acrodef{LBFN}[LBFN]{Linear Beamforming Network}
\acrodef{LLF}[LLF]{Log-Likelihood function}
\acrodef{LMD}[LMD]{Linearly Modulated Digital}
\acrodef{LoS}[LoS]{Line-of-Sight}
\newacro{MAP}{Maximum A Posteriori}
\acrodef{MIMO}[MIMO]{Multiple-Input Multiple-Output}
\newacro{ML}{Maximum Likelihood}
\newacro{MMSE}{Minimum Mean Squared Error}
\acrodef{MMV}[MMV]{Multiple Measurement Vector}
\acrodef{mmWave}[mmWave]{millimeter wave}
\acrodef{MRC}[MRC]{Maximum Ratio Combining}
\acrodef{MRT}[MRT]{Maximum Ratio Trasmission}
\newacro{MSE}{Mean Squared Error}
\acrodef{MUSIC}[MUSIC]{MUltiple SIgnal Classification}
\acrodef{NLOS}[NLOS]{Non Line-of-sight}
\acrodef{NMSE}{Normalized Mean Squared Error}
\acrodef{OFDM}[OFDM]{Orthogonal Frequency-Division Multiplexing}
\acrodef{OFDMA}[OFDMA]{Orthogonal Frequency-Division Multiple Access}
\acrodef{OMP}[OMP]{Orthogonal Matching Pursuit}
\acrodef{OSMP}[OSMP]{Orthogonal Subspace Matching Pursuit}
\acrodef{PA}[PA]{Power Amplifier}
\acrodef{PS}[PS]{Phase Shifter}
\acrodef{PSK}[PSK]{Phase-Shift Keying}
\acrodef{QAM}[QAM]{Quadrature Amplitude Modulation}
\acrodef{RF}[RF]{Radio Frequency}
\acrodef{RFC}[RFC]{Rayleigh Fading Channel}
\acrodef{SDMA}[SDMA]{space-division multiple access}
\acrodef{SER}[SER]{Symbol Error Rate}
\acrodef{SIC}[SIC]{Successive Interference Cancellation}
\acrodef{SR}[SR]{Sideband Radiation}
\acrodef{SINR}[SINR]{Signal-to-Interference-plus-Noise Ratio}
\acrodef{SLL}[SLL]{Side-Lobe Level}
\acrodef{SOCP}[SOCP]{Second-Order Cone Program}
\acrodef{SOMP}[SOMP]{Simultaneous-Orthogonal Matching Pursuit}
\acrodef{SPDT}[SPDT]{Single-Pole-Double-Throw}
\acrodef{SPST}[SPST]{Single-Pole-Single-Throw}
\acrodef{SR}[SR]{Sideband Radiation}
\acrodef{SS}[SS]{Spatial Smoothing}
\acrodef{SNR}[SNR]{Signal-to-Noise Ratio}
\acrodef{SUS}{successive user selection}
\newacro{TDD}{Time-Division Duplex}
\acrodef{TM}[TM]{Time Modulation}
\acrodef{TMA}[TMA]{Time-Modulated Array}
\acrodef{ULA}[ULA]{Uniform Linear Array}
\acrodef{UPA}[UPA]{Uniform Planar Array}
\acrodef{VGA}[VGA]{Variable Gain Amplifier}
\acrodef{VPS}[VPS]{Variable Phase Shifter}
\acrodef{VR}[VR]{visibility regions}
\acrodef{XL}[XL]{extra-large}
\acrodef{ZF}[ZF]{Zero-Forcing}

\makeatletter
\def\blfootnote{\xdef\@thefnmark{}\@footnotetext}
\makeatother

\begin{document}

\title{\LARGE{Low-Complexity Distance-Based Scheduling\\ for Multi-User XL-MIMO Systems}}

\author{José~P.~Gonz\'alez-Coma,~\IEEEmembership{Member,~IEEE,}
	 F. Javier~L\'opez-Mart\'inez,~\IEEEmembership{Senior Member,~IEEE,}
	and~Luis~Castedo,~\IEEEmembership{Senior Member,~IEEE,}}

\maketitle
\begin{abstract}
\blfootnote{\noindent The work of F.J. L\'opez-Mart\'inez was funded by Junta de Andalucia and the European Fund for Regional Development FEDER (project P18-RT-3175). The work of J.P. Gonz\'alez-Coma and L. Castedo has been funded by the Xunta de Galicia (ED431G2019/01), the Agencia Estatal de Investigación of Spain (TEC2016-75067-C4-1-R) and ERDF funds of the EU (AEI/FEDER, UE).} 
\blfootnote{\noindent J.P. Gonz\'alez-Coma is with Defense University Center at the Spanish Naval Academy, Marín 36920, Spain. F. J. L\'opez-Mart\'inez is with Departmento de Ingenier\'ia de Comunicaciones, Universidad de M\'alaga - Campus de Excelencia Internacional Andaluc\'ia Tech., M\'alaga 29071, Spain. L. Castedo is with Department of Computer Engineering \& CITIC Research Center, University of A Coru\~na,  A Coru\~na 15001, Spain.}
\blfootnote{This work has been submitted to the IEEE for possible publication. Copyright may be transferred without notice, after which this version may no longer be accessible.}
We introduce \ac*{DBS}, a new technique for user selection in downlink multi-user communications with extra-large (XL) antenna arrays. DBS categorizes users according to their \emph{equivalent distance} to the antenna array. Such categorization effectively accounts for inter-user interference while largely reducing the computational burden. 
Results show that (\emph{i}) DBS achieves the same performance as the reference zero-forcing beamforming scheme with a lower complexity; (\emph{ii}) a simplified version of DBS  achieves a similar performance when realistic spherical-wavefront (SW) propagation features are considered; (\emph{iii}) SW propagation brings additional degrees of freedom, which allows for increasing the number of served users. 
\end{abstract}
\IEEEpeerreviewmaketitle
\vspace{-2mm}
\vspace{-3mm}
\section{Introduction}
The successful deployment of multi-user \ac{MIMO} technology in the context of the 5G standard has been made possible by the massive \ac{MIMO} concept \cite{Marzetta2010}. As the commercial validation of massive \ac{MIMO} is a key milestone in the roadmap of multi-user \ac{MIMO}, the next steps move towards pushing the number of antennas and served users to a 10$\times$increase \cite{Emil2019}. In this situation, the size of the \ac{XL} antenna arrays becomes comparable to the user distances and the conventional far-field assumption no longer holds. Instead, non-stationary channel features need to be considered \cite{Carvalho2020}, which affect the scaling laws of the \ac{SNR} as the number of antennas grows \cite{LuZe20}. 

Capacity-achieving schemes in multi-user \ac{MIMO} based on \ac{DPC} have a prohibitive complexity. Thus, the use of sub-optimal linear strategies based on \ac{ZF} precoding to remove inter-user interference is usually preferred. \ac{ZF}-beamforming (ZFBF) is known to have nearly as good performance as \ac{DPC}, provided that a set of semi-orthogonal users are available for transmission \cite{YoGo06}. However, since the computational complexity of linear precoders grows with the system dimensions, i.e., antennas and users \cite{Muller2016}, the problem of low-complexity user scheduling in \ac{XL}-\ac{MIMO} needs to be better examined. Besides, the interference behavior in \ac{XL}-\ac{MIMO} heavily depends on the user distances to the array elements, and interference can be incorrectly estimated when users are close to the antenna array \cite{Carvalho2020}. 

The use of simple linear precoding techniques in the \ac{XL}-\ac{MIMO} regime was recently evaluated in \cite{Ali2019}, showing the impact of spatial non-stationarity on the \ac{DL} performance. The notion of \ac{VR} may partially alleviate the computational burden associated to \ac{ZF} operation and \ac{MRT} through antenna selection \cite{Marinello2020}, although the large array dimension still poses important challenges from a complexity viewpoint. Very recently, two low-complexity precoding schemes were proposed in \cite{Ribeiro2021} by a proper user grouping in the elevation domain. However, since both methods make use of a plane-wave (PW) approximation, their performance is degraded {compared to the reference ZFBF for users closer to the \ac{BS}}.

In this paper, we present two low-complexity techniques for \ac{DL} user scheduling in \ac{XL}-\ac{MIMO} systems. Based on a novel definition of \emph{equivalent distance} that accounts for the interference level, the user distance to the center of the array, and the number of antennas, we propose two schemes on which users are selected for transmission based on their {equivalent distance} to the \ac{BS}; hence, we refer to these techniques as \ac{DBS}. We show that \ac{DBS} achieves the same performance as conventional ZFBF with a much lower complexity, while effectively capturing the spherical wavefront (SW) propagation experienced by users in the near-field region of the antenna array. We also see that a simplified-\ac{DBS} scheme allows for an even lower complexity at the expense of a moderate performance degradation.

\textit{Notation:} Throughout the article, lower-case bold letters denote vectors; the symbol $\sim$ reads as \emph{statistically distributed as}; $\left ( \cdot \right )^\Tr$ and $\left ( \cdot \right )^\He$ denote the transpose and Hermitian transpose operations, respectively; ${\bf{1}}$ is the all-one vector; $\mathcal{N}_\mathbb{C}(0,\sigma^2 )$ is the zero-mean circularly symmetric Gaussian distribution with variance $\sigma^2$; $\|\cdot\|$ is the Euclidean norm; $\mathbb{C}^M$ is the $M$-dimensional complex vector space.

\vspace{-4mm}
\section{System model}
\label{sec:model}
We consider an \ac{XL}-\ac{MIMO} setup where the \ac{BS} deploys an extra-large antenna array with $M\gg 1$ elements, and communicates with $K$ single-antenna users. 
Without loss of generality, we assume a \ac{ULA} centered at the origin $\rm O$ and deployed along the ordinate axis. The position of the $k$-th user is then determined by the distance to the antenna array center, $r_k$, and the angle, $\theta_k$, formed by the line connecting user $k$ to $\rm O$ and the abscissa axis. Therefore, the array response vector ${\bf{a}}_{\rm k}\in\mathbb{C}^M$ reads as
\begin{equation}\vspace{-1mm}
	{\bf{a}}_{\rm k}=[a_1(r_k,\theta_k),a_2(r_k,\theta_k),\ldots,a_M(r_k,\theta_k))]^\Tr,\vspace{-1mm}
\end{equation}
where 
\begin{equation}\vspace{-1mm}
a_m(r_k,\theta_k)=\tfrac{\sqrt{\beta_0}}{r_{k,m}}\eul^{-\imj\frac{2\pi}{\lambda}r_{k,m}},
\label{eq:antenaElemSW}\vspace{-1mm}
\end{equation}
is the $m$-th element of ${\bf{a}}_{\rm k}$, $\beta_0$ denotes the channel power at the reference distance $r_{\rm ref}=1$m, and $\lambda$ is the signal wavelength \cite{LuZe20}. Finally, $r_{k,m}$ stands for the distance between the $k$-th user and the $m$-th element of the antenna array as
\begin{align}\vspace{-1mm}
	r_{k,m}=r_k\sqrt{1-2m{d}_k\sin\theta_k+{d}_k^2m^2},\,m\in\left[-\tfrac{M}{2},\tfrac{M}{2}\right],   
	\label{eq:antennaDistance}\vspace{-1mm}
\end{align}
with $d_k=\frac{d}{r_k}$, and $d$ is the separation between two consecutive antenna elements. As in \cite{LuZe20,Zhou2015}, we assume that the \ac{LoS} component dominates the channel vector response. 

Prior to transmission, the data symbols $s_k\sim\mathcal{N}_\mathbb{C}(0,1)$, $k=1,\ldots,K$ are precoded using the precoding vectors ${\bf{f}}_{\rm k}\in\mathbb{C}^M$, with $\|{\bf{f}}_{\rm k}\|=1$. The transmitted signal ${\bf{x}}_{\rm k}$ is then a linear combination of the precoded symbols, i.e., ${\bf{x}}_{\rm k}=\sum_k^Kp_k{\bf{f}}_{\rm k}^\He s_k$, where  $p_k$ are the power allocation scale factors such that $\sum_{k=1}^Kp_k\leq P_\text{TX}$, with $P_\text{TX}$ the \ac{BS} transmit power. For user $k$, the received signal $y_k$ is affected by the channel vector and the \ac{AWGN} $n_k\sim\mathcal{N}_\mathbb{C}(0,\sigma^2_w)$, as
\begin{equation}
\label{eq4}\vspace{-1mm}
	y_k=p_ks_k{\bf{f}}_{\rm k}^\He{\bf{a}}_{\rm k}+\sum_{j\neq k}p_js_j{\bf{f}}_{\rm j}^\He{\bf{a}}_{\rm k}+w_k,
\vspace{-1mm}
\end{equation}
where the first term in \eqref{eq4} is the desired signal, and the second one represents the inter-user interference. Given the former expression, the achievable sum-rate is defined as \cite{YoGo06}
\begin{align}
\label{eq:sumRate}\vspace{-1mm}
R=\sum_{k=1}^{K}R_k=\sum_{k=1}^{K}\log_2\Bigg(1+\underbrace{\tfrac{|p_k{\bf{f}}_{\rm k}^\He{\bf{a}}_{\rm k}|^2}{\sigma^2_w+\sum_{j\neq k}|p_j{\bf{f}}_{\rm j}^\He{\bf{a}}_{\rm k}|^2}}_{\rm{SINR}_k}\Bigg),\vspace{-1mm}
\end{align}
where $\text{SINR}_k$ accounts for the $k$-th user \ac{SINR}. Our aim is to maximize the sum-rate in \eqref{eq:sumRate} subject to the power constraint, i.e.,
\begin{equation}\vspace{-1mm}
	\argmax_{\{{\bf{f}}_{\rm k},p_k\}_{k=1}^K} R\quad\text{s.t.}\quad\sum_{k=1}^Kp_k\leq P_\text{TX}.
	\label{eq:problemForm}\vspace{-.75mm}
\end{equation}

We note from \eqref{eq:sumRate} that the achievable sum-rate is limited by the inter-user interference. However, the behavior of such interference severely changes with $r_{k,m}$ when near-field propagation effects are accounted for: e.g., the interference caused by users close to the \ac{BS} is underestimated when assuming the PW model instead of \eqref{eq:antenaElemSW}. Conversely, as the spatial signature for each user depends on both their distances and their angles, when users $j$ and $k$ have similar angular values $\theta_j\approx\theta_k$ but different distances $r_j\neq r_k$, then the actual interference under the SW model is lower than that predicted by the PW assumption. With all the above considerations, a proper choice of the set of users $\mathcal{S}\subseteq\{1,\ldots,K\}$  served by the \ac{BS} is crucial for the good functioning of the precoder and power allocation strategies. Hence, we split the problem in \eqref{eq:problemForm} into two sub-problems: \emph{(i)} the selection of a user set $\mathcal{S}$ served by the \ac{BS} according to the instantaneous channel conditions; and \emph{(ii)} the precoder design and power allocation for the selected set $\mathcal{S}$. 

\vspace{-3mm}
\section{User scheduling in XL-MIMO}
\label{sec:scheduling}
State-of-the-art user schedulers for massive \ac{MIMO} setups employ the similarity between the channel vectors or the channel correlation matrices to perform user selection. These strategies become prohibitively complex in the \ac{XL}-\ac{MIMO} regime \cite{AdNaAhCa13}. Besides, they often require a combinatorial search that is unfeasible for configurations with large $K$ and $M$ \cite{Muller2016}. Greedy approximations \cite{YoGo06,GuUtDi09} reduce the number of combinations to check. This {allows} practical \ac{ZF} schemes to achieve the capacity of \ac{DPC} under the assumptions:  \emph{(i)} channel vectors with independent Gaussian entries and asymptotically large $K$, and \emph{(ii)}  PW model and sufficiently large $M$. Throughout this section, we propose a novel scheduling technique that largely alleviates the still large computational cost of conventional greedy schemes, while taking advantage of the specific features of \ac{XL}-\ac{MIMO} channels. 

\vspace{-3mm}
\subsection{Distance-Based Scheduling}
We consider a greedy approach to joint scheduling and precoding inspired by superposition coding, where users are decoded according to their \ac{SNR} levels. In our multi-user setup, we use the \ac{SINR} definition in \eqref{eq:sumRate} so that users with larger \acp{SINR} are assigned higher priorities and, correspondingly, a larger allocation of the transmit power $P_\text{TX}$. In order to incorporate the interference caused by users enjoying higher priority levels, we base our scheduling policy on the concept of \emph{equivalent distance}.

 Let us consider the $n$-th iteration of the greedy procedure; the equivalent distance is defined as
\begin{equation}
r_{\text{eq},k}^{(n)}=r_k
\Bigg(1-\frac{r_k^2}{M}\sum_{j\in\mathcal{S}^{(n)}}|{\bf{f}}^{\He,(n)}_{\rm j}{\bf{a}}_{\rm k}|^2\Bigg)^{-1/2},
\label{eq:update_distance}
\end{equation}
where $\mathcal{S}^{(n)}\subseteq\{1,\ldots,K\}$ is the sequence containing the $n$ served users, and ${\bf{f}}_{\rm j}^{(n)}$ is the precoder associated to user $j$, with $j\in\mathcal{S}^{(n)}$. In the initialization step, the equivalent distance $r_{\text{eq},k}^{(0)}$ is equal to the distance between the user and the \ac{BS}, i.e., $r_{\text{eq},k}^{(0)}=r_k$. In \eqref{eq:update_distance}, the equivalent distance is updated according to the level of interference caused by the scheduled users $\mathcal{S}^{(n)}$, the user distance to the \ac{BS} $r_k$, and the number of antennas $M$. We see that for $\tfrac{r_k^2}{M}\ll 1$, the equivalent distance barely deviates from $r_k$ when including the interference caused by higher priority users. Conversely, when $\tfrac{r_k^2}{M}\gg 1$ inter-user interference strongly affects the computation of the equivalent distance. For a given $M$, these different behaviors are related to the distances between the users and the \ac{BS}. The equivalent distance in \eqref{eq:update_distance} is used to assign higher priority levels to users close to the \ac{BS}, as the impact of the inter-user interference is smaller than for users further away; therefore, we refer to this strategy as \textit{Distance-Based scheduling} (DBS). In practice, the equivalent distance for user $k$ can be interpreted as an approximation to its $\text{SINR}_k$.

The set of precoders ${\bf{F}}(\mathcal{S}^{(n)})=[{\bf{f}}^{(n)}_{\rm k_1},\ldots,{\bf{f}}^{(n)}_{\rm k_n}]$, with $k_1,\ldots,k_n\in\mathcal{S}^{(n)}$ is used to compute the equivalent distances \eqref{eq:update_distance}, as the amount of interference depends on this choice. Considering the characteristics of the \ac{XL}-\ac{MIMO} setup with regard to the available spatial degrees of freedom and \ac{SNR} regime, it is reasonable to apply \ac{ZF} precoding at each iteration
\begin{equation}
\overline{{\bf{F}}}(\mathcal{S}^{(n)})=\left[{\bf{A}}(\mathcal{S}^{(n)})^\He\left({\bf{A}}(\mathcal{S}^{(n)}){\bf{A}}(\mathcal{S}^{(n)})^\He\right)^{-1}\right],
\label{eq:ZFprecoders}
\end{equation}
where ${\bf{A}}(\mathcal{S}^{(n)})=[{\bf{a}}^{(n)}_{\rm k_1},\ldots,{\bf{a}}^{(n)}_{\rm k_n}]^\Tr$ stacks the channel vectors corresponding to the sequence $\mathcal{S}^{(n)}$. To obtain the set of precoders ${\bf{F}}(\mathcal{S}^{(n)})$, we normalize the columns of $\overline{{\bf{F}}}(\mathcal{S}^{(n)})$. The equivalent channel gains $|{\bf{f}}_{\rm k}^\He{\bf{a}}_{\rm k}|^2,\,k\in\mathcal{S}^{(n)}$ are then employed to compute the power allocation weights for the set of scheduled users ${\bf{p}}^{(n)}(\mathcal{S}^{(n)})\in\mathbb{R}^n$ using waterfilling. 

\begin{algorithm}[t]
	\vspace{-.5mm}
	\caption{Distance-Based Scheduling}\label{alg:Scheduler}
	\begin{algorithmic}[1]
	\small
		\STATE  $n \gets 0$, $\mathcal{S}^{(0)}\gets\emptyset$, $\mathcal{K}\gets\{1,\ldots,K\}$
		\STATE $r_{\text{eq},k}^{(0)} \gets$ initialization 
		\REPEAT
		\REPEAT
		\STATE $k\gets \min_{i\in\mathcal{K}}r_{\text{eq},i}^{(n)}$
		\STATE $r_{\text{eq},k}^{(n)}\gets$ update using \eqref{eq:update_distance} \label{sch:update}
		\STATE $q\gets \min_{i\in\mathcal{K}}r_{\text{eq},i}^{(n)}$
		\UNTIL{$q=k$}
		\STATE  $n \gets n+1$
		\STATE $\mathcal{S}^{(n)}\gets\mathcal{S}^{(n-1)}\cup\{k\}$
		\STATE $\overline{{\bf{F}}}(\mathcal{S}^{(n)})\gets$ Compute ZF precoders \eqref{eq:ZFprecoders}
		\STATE ${\bf{F}}(\mathcal{S}^{(n)})\gets$ Normalize $\overline{{\bf{F}}}(\mathcal{S}^{(n)})$ columns 
		\STATE ${\bf{p}}^{(n)}(\mathcal{S}^{(n)})\gets$ Power allocation using waterfilling
		\STATE $\mathcal{K}\gets\mathcal{K}\setminus\{k\}$
		\UNTIL{$\mathcal{K}=\emptyset$ or stopping criterion}
	\end{algorithmic}
	\vspace{-.5mm}
\end{algorithm}

The proposed procedure for the \ac{DBS} algorithm is summarized in Alg. \ref{alg:Scheduler}: users start with their corresponding distances, and the algorithm seeks for the closest one to the \ac{BS}. As new users are allocated, the equivalent distances are updated according to \eqref{eq:update_distance}. If the user ordering does not change after the update, the closest user is selected as the candidate. Otherwise, the search continues with the following closest user. When the candidate user is selected, \ac{ZF} precoders are computed according to the sequence $\mathcal{S}^{(n)}$, and then normalized to obtain the equivalent channel gains. Finally, these gains are used to decide the power distribution among users, ${\bf{p}}^{(n)}(\mathcal{S}^{(n)})$, with ${\bf{1}}^\Tr{\bf{p}}^{(n)}(\mathcal{S}^{(n)})=P_\text{TX}$. The algorithm ends after some stopping criterion is met like, for instance, a reduction on the achievable sum-rate or the maximum equivalent distance.

One interesting feature of \ac{DBS} relies on the complexity reduction compared to the baseline greedy methods \cite{YoGo06,GuUtDi09}, as we only evaluate the users within a certain distance range at each iteration of Alg. \ref{alg:Scheduler}. This is especially noteworthy as the number of users $K$ grows and becomes comparable to $M$. Moreover, as we employ equivalent distances to determine the user ordering, most comparisons only involve scalars. Therefore, it is not only the number of comparisons that we reduce in DBS, but also the computational cost of making each of them. As we will later see, the relevant complexity reduction offered by \ac{DBS} does not come at the price of a performance degradation compared to the ZFBF solutions \cite{YoGo06}.
\vspace{-3mm}
\subsection{Simplified DBS}

DBS algorithm allows to decrease the complexity associated to user selection; however, a moderately costly operation is necessary to update the equivalent distance in line \ref{sch:update} of the algorithm. We next evaluate an alternative to further reduce complexity at the expense of a certain performance loss. 

From \eqref{eq:update_distance}, it is clear that users with reduced $r_k$ are likely to be selected in the scheduling process, as the impact of the interference on the equivalent distance is proportional to $r_k^2$. Motivated by this observation, and noticing that an expensive step for the proposed \ac{DBS} method is the distance update, we propose to use a naive approach that avoids this step. Therefore, the simplified \ac{DBS} scheme, that we will refer to as \ac{DBS}-s, will simply iterate using the distances $r_k$, regardless of the inter-user interference suffered by users with smaller priorities. Again, this approximation includes new users until a certain stopping criterion is met. 

Note that, although the speed of this method is superior compared to regular \ac{DBS}, the achievable performance of \ac{DBS}-s strongly depends on the particular characteristics of the scenario. Consider, for example, a setup where users $1$ and $2$ are close to each other and satisfy $r_1<r_2\leq r_k$, $\forall k\in\{3,\ldots,K\}$. As the equivalent distance of user $2$ is not updated, the \ac{DBS}-s algorithm would only serve user $1$. Hence, it is expected that the \ac{DBS}-s scheme may incur in a performance loss, both in terms of served users and achievable rates. This will be assessed further in Section \ref{sec:numerical}.
\vspace{-2mm}
\section{Numerical results}
\label{sec:numerical}

In this section, we provide the results of simulation experiments conducted to assess the advantages of the proposed scheduling methods. Table \ref{tab:Sim1} shows the simulation parameter settings. The distance range considered is set to roughly twice the \emph{critical distance} $r_\text{cri}=9Md$ in \cite{LuZe20}, which is the effective distance of the bound that separates the  near-field and far-field regions. Since user distances within the coverage area are uniformly distributed, the proportion of users located in the near-field and far-field region is approximately $50\%$.

For benchmarking purposes, we compare the performance of the \ac{DBS} and \ac{DBS}-s methods with the classical linear precoding schemes considered in \cite{Ali2019}. Specifically, we consider (\textit{i}) the \ac{MRT} precoder design followed by a waterfilling power allocation, and (\textit{ii}) the greedy approach in \cite{GuUtDi09} for {ZFBF} with \ac{SUS}. The stopping criterion for the greedy approaches is a reduction on the achievable sum-rate. The SW propagation model in Section \ref{sec:model} is used, although reference results for the (incorrect) PW model associated to the far-field assumption are also included using dashed lines.

\begin{table}[t]
	\centering
	\caption{{Simulation parameter settings.}}\label{tab:Sim1}
	\setlength{\tabcolsep}{5pt}
	\def\arraystretch{1.5}
	\begin{tabular}{|l|l|}
		\hline		
		{\textbf{Parameter}} & {\textbf{Value}}  \\ \hline\hline
		Number of users & $K=1000$ \\ \hline
		Distance between antennas & $d=0.0628$ m \\\hline
		Transmit SNR ($\beta_0/\sigma^2_w$)& $\{0,5,10,15,20,25\}$ dB\\\hline
		Distance range &$ [40, 2r_\text{cri}-40]$ m\\\hline
		Angular range &$[-\frac{\pi}{4} \frac{\pi}{4}]$\\\hline
		Channel realizations & $1000$\\
		\hline
	\end{tabular}
\end{table}

\begin{figure}[t]
	\centering
	\includegraphics[width=.84\columnwidth]{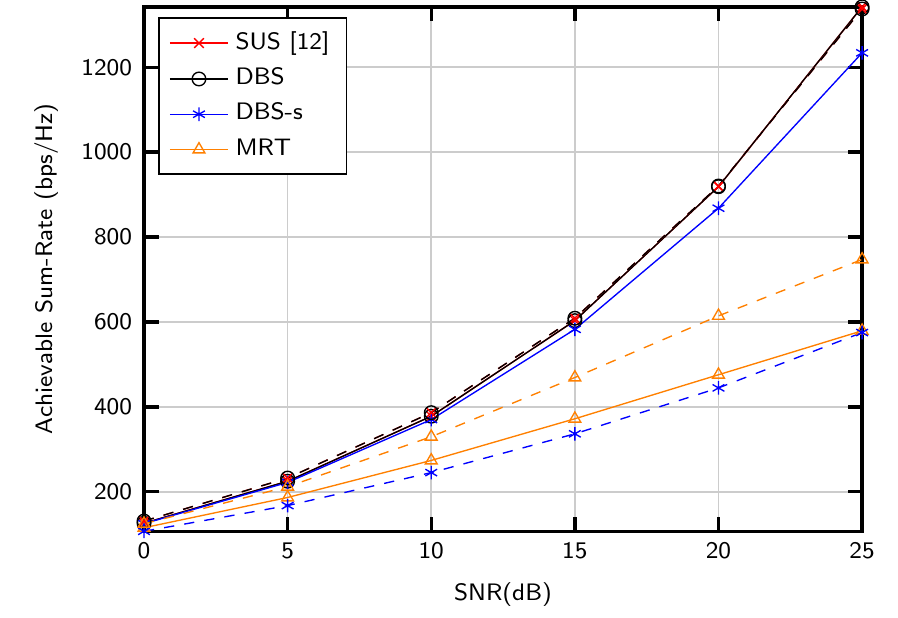}
	\caption{Achievable sum rate vs. SNR (dB) for $M=1000$ antennas. Parameter values are given in Table \ref{tab:Sim1}. Solid/dashed lines correspond to the SW/PW models, respectively.} 
	\label{fig:SNR}
\end{figure}

Fig. \ref{fig:SNR} shows the achievable sum-rate $R$ as a function of the transmit \ac{SNR} for all the schemes under evaluation. The first important observation is that the results obtained for the \ac{DBS} and \ac{SUS} schemes are equivalent regardless of the use of the SW or PW models, although for the latter model the achievable rate is slightly overestimated. This confirms that despite of using the \emph{equivalent distance} as an approximation to the equivalent channel gains, there is no performance loss compared to the reference \ac{SUS}. Remarkably, this will occur along all the ensuing investigated configurations. The performance of the \ac{MRT} scheme notably degrades under SW propagation conditions. The opposite behavior is observed for the \ac{DBS}-s scheme, for which the achievable rate is rather close to the reference schemes \ac{DBS}/\ac{SUS}. Hence, the \ac{DBS} scheme can be seen as a simplified version of the \ac{SUS} scheme but still offering, with lesser complexity, the best performance among the linear precoding schemes considered. This is confirmed by the execution times included in Table \ref{tab:execution_time} which shows how \ac{DBS} allows for a noticeable complexity reduction (around $80\%$) compared to \ac{SUS}. For the \ac{DBS}-s scheme, such reduction can even go well beyond $90\%$ at the expense of a minor performance degradation.

\begin{table}[t] 
	\centering
	\setlength{\tabcolsep}{5pt}
	\def\arraystretch{1.5}
	\caption{Execution time (ms)}
	\begin{tabular}{|c|c|c|c|c|c|c|}
		\hline
		Method/SNR(dB)  & \textbf{0} & \textbf{5}  & \textbf{10} & \textbf{15} & \textbf{20} & \textbf{25} \\
		\hline\hline
		\textbf{SUS} \cite{GuUtDi09}		& 3.65	& 5.35	&  7.46 & 10.18	& 13.25	&18.14\\
		\textbf{DBS}					& 0.35	& 0.57	&  0.94 & 1.50	& 2.33	&3.72\\
		\textbf{DBS-s}					& 0.16	& 0.30	&  0.50 & 0.76	& 1.15	&1.70\\
		\hline
	\end{tabular}
	\label{tab:execution_time}
\end{table}

\begin{figure}[t]
	\centering
	\includegraphics[width=.84\columnwidth]{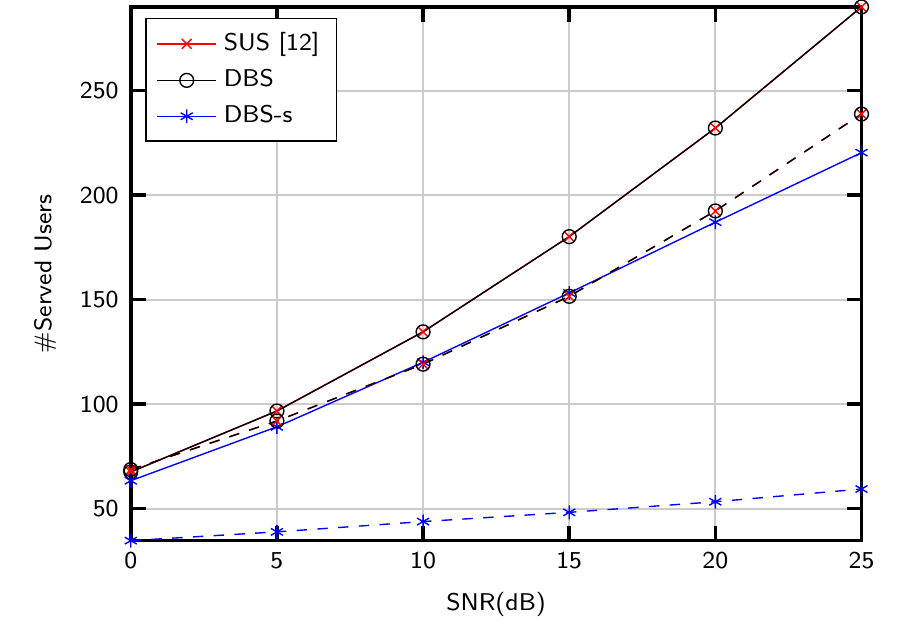}
	\caption{Number of served users vs. SNR (dB) for $M=1000$ antennas. Parameter values are given in Table \ref{tab:Sim1}. Solid/dashed lines correspond to the SW/PW models, respectively.} 
	\label{fig:servedUsers}
\end{figure}

Fig. \ref{fig:servedUsers} shows the number of served users as a function of the transmit \ac{SNR}. As in the previous figure, the performance of \ac{SUS} and \ac{DBS} schemes is perfectly coincident. We see that a noticeable increase in served users is achieved when the SW propagation is accounted for, which is explained by the different interference behaviors exhibited by users closer to the antenna arrays. This provides more flexibility in the user selection, and allows to schedule a number of users larger than under the PW assumption. We also observe that while the performance of \ac{DBS}-s seems poor under the PW assumption, it improves dramatically when the SW model is considered.

\begin{figure}[t]
	\centering
	\includegraphics[width=.84\columnwidth]{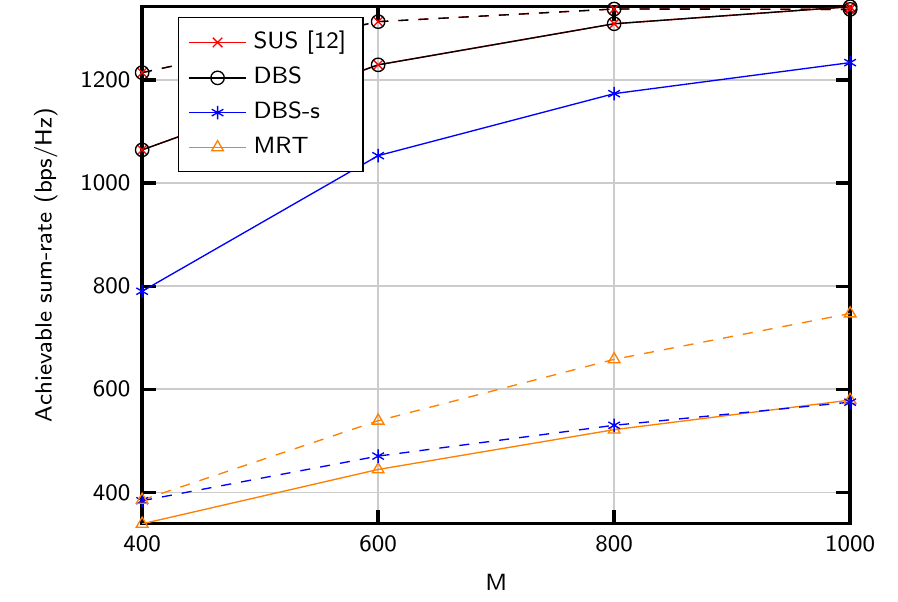}
	\caption{Achievable sum rate vs. number of antennas $M$ for SNR$=25$ dB. Parameter values are given in Table \ref{tab:Sim1}. Solid/dashed lines correspond to the SW/PW models, respectively.} 
	\label{fig:antenas}
\end{figure}

Finally, the performance in terms of the number of antennas is evaluated in Fig. \ref{fig:antenas}. As justified in Section \ref{sec:scheduling}, the far-field approximation underestimates inter-user interference for users close to the \ac{BS}, thus providing exceedingly optimistic performance results, especially when the number of antennas is small compared to the number of users. This observation is also supported by the results obtained with the simple \ac{MRT} design. Conversely, the achievable sum-rates for the \ac{DBS}-s scheme largely improves again under SW propagation conditions, so that including users close to the \ac{BS} is key to obtain good performance results. This confirms the important role of the equivalent distance parameter accounting for the near-field propagation: for instance, the amount of inter-user interference between users with similar angular values is small if the distances are different. 
\vspace{-2mm}
\section{Conclusion}
\label{sec:conc}
The use of \ac{DBS} schemes for user selection in \ac{DL} \ac{XL}-\ac{MIMO} systems reduces the complexity of state-of-the-art methods. The performance of \ac{DBS} is the same as that of the reference ZFBF approach. The complexity of \ac{DBS} may be further reduced with a minor performance degradation under SW propagation. As a final remark, we pose that the optimality conditions for ZFBF may not hold in the \ac{XL}-\ac{MIMO} regime, as shown in the Appendix. This suggests that the development of capacity-approaching precoding techniques for \ac{XL}-\ac{MIMO} requires further investigation, although we conjecture that ZFBF (and hence, \ac{DBS}) still can achieve the best performance among linear precoding schemes. 
\section*{Appendix: On the optimality of ZFBF}
Linear precoding achieves the performance of \ac{DPC} under certain assumptions \cite{YoGo06}. In \ac{XL}-\ac{MIMO}, however, this conclusion does not apply. In the following, we prove our statement constructively using a counter example. 

Let us define a metric to characterize the power distribution among the transmit antenna elements. Such metric is given by the quotient between the normalized channel vector norm $\frac{1}{M}\|\B{a}_{\rm k}\|^2=\frac{\beta_0\varphi}{Mr_kd\cos(\theta_k)}$ from \cite{LuZe20} and  the maximum per-antenna-element power, $|a_n(r_k,\theta_k)|^2$, i.e.,
\begin{align}
c(r_k,\theta_k)&=\tfrac{r_{k,n}^2\varphi}{Mr_kd\cos(\theta_k)},
\label{eq:equity}	
\end{align}	
where $n\in\{-\frac{M}{2},\ldots,\frac{M}{2}\}$ is the index corresponding to the antenna array element closest to user $k$, and $\varphi\in[0,\pi]$ is
\begin{equation*}
\varphi=\atan\left({\tfrac{Md-2r_k\sin(\theta_k)}{2r_k\cos(\theta_k)}}\right)+\atan\left(\tfrac{Md+2r_k\sin(\theta_k)}{2r_k\cos(\theta_k)}\right).
\end{equation*}  

Next, consider users $k$ and $j$ and their associated distances and angles, i.e., $r_k$ and $r_j$, and $\theta_k$ and $\theta_j$, respectively. We now introduce the interference experienced by user $j$ if user $k$ employs the \ac{MRT} precoder  $\B{f}_{\rm k}=\frac{1}{\|\B{a}_{\rm k}\|}\B{a}^\He_{\rm k}$. This leads to
\begin{align}
|\B{f}_{\rm k}^\He\B{a}_{\rm j}|=\left|\sum_{m=-M/2}^{M/2}\tfrac{\beta_0}{\|\B{a}_{\rm k}\|r_{m,i}r_{m,j}}\eul^{-\imj\frac{2\pi}{\lambda}(r_{m,j}-r_{m,i})}\right|,
\label{eq:interf}
\end{align}
with $r_{m,k}$ defined in \eqref{eq:antennaDistance}. Without loss of generality, we assume that $r_j>r_k$, and consider $\theta_k=0$ for ease of exposition. Now, for $\frac{r_k^2}{M}\ll 1$, equation \eqref{eq:equity} shows that the transmit power concentrates over a reduced number of antennas $M^\prime\ll M$  close to the center of the array. Hence, for the significant antenna elements $m\in\{-M^\prime/2,\ldots,M^\prime/2\}$, the far-field assumption approximately holds and we have $a_m(r_k,\theta_k)\approx\tfrac{\sqrt{\beta_0}}{r_{k}}\eul^{-\imj\frac{2\pi}{\lambda}r_{k}}$ and
$a_m(r_j,\theta_j)\approx\tfrac{\sqrt{\beta_0}}{r_{j}}\eul^{-\imj\frac{2\pi}{\lambda}(r_j-md\sin(\theta_j))}$. We rewrite \eqref{eq:interf} as
\begin{align}
|\B{f}_{\rm k}^\He\B{a}_{\rm j}|&\approx\tfrac{\beta_0}{\|\B{a}_{\rm k}\|r_kr_j}\left|\sum_{m=-M^\prime/2}^{M^\prime/2}\eul^{-\imj\frac{2\pi}{\lambda}(r_j-md\sin(\theta_j)-r_k)}\right|\nonumber\\
&=\tfrac{\beta_0}{\|\B{a}_{\rm k}\|r_kr_j}\left|\tfrac{\sin\left(\frac{\pi}{\lambda}dM^\prime\sin(\theta_j)\right)}{\sin\left(\frac{\pi}{\lambda}d\sin(\theta_j)\right)}\right|=i(\theta_j),
\label{eq:interfApp}
\end{align}
where $i(\theta_j)$ accounts for the interference incident through angle $\theta_j$. 

Next, we evaluate the probability of finding a set of semi-orthogonal users in the angular direction $\theta_j$. We introduce the interference threshold $\alpha$ and define the probability function $P\{i(\theta_j)<\alpha\}$. To determine this probability, we define $\frac{M^\prime-1}{2}$ partitions $\mathcal{A}_q$ of the interval $[0,\frac{\pi}{2}]$, $\mathcal{A}_q=\left[l_{q,L},l_{q,U}\right]$,  
whose boundaries satisfy $i(l_{q,L})=i(l_{q,U})=0$. This condition is met for $l_{q,L}=\arcsin\left(\frac{\lambda q}{dM^\prime}\right)$, $l_{q,U}=\arcsin\left(\frac{\lambda (q+1)}{dM^\prime}\right)$, and $q\in\{0,(M^\prime-3)/2\}$. 
Based on these partitions, we determine $P\{i(\theta_j)<\alpha\}$ as
\begin{align}
\label{eq:prob}
&P\{i(\theta_j)<\alpha\}=2\sum_{q=0}^{\frac{M^\prime-1}{2}}P\{i(\theta_j)<\alpha\,\cap\,\theta_j\in\mathcal{A}_q\}\\
&\leq 2\sum_{q=0}^{\frac{M^\prime-1}{2}}P\left\{\left|\sin\left(\tfrac{\pi}{\lambda}dM^\prime\sin(\theta_j)\right)\right|<\sin\left(\tfrac{\pi q}{M^\prime}\right)\alpha^\prime\,\cap\,\theta_j\in\mathcal{A}_q\right\}\nonumber
\end{align}
where $\alpha^\prime=r_kr_j\alpha\|\B{a}_{\rm k}\|\beta_0^{-1}$. The inequality results from bounding the denominator in \eqref{eq:interfApp} as $\sin\left(\frac{\pi}{\lambda}d\sin(\theta_j)\right)\leq\sin\left(\frac{\pi q}{M^\prime}\right)$ for $\theta_j\in\mathcal{A}_q$; equality holds if $M^\prime\rightarrow\infty$. To compute \eqref{eq:prob}, observe that there are two angles $\phi_{q,L},\phi_{q,U}\in\mathcal{A}_q$ fulfilling   $\left|\sin\left(\frac{\pi}{\lambda}dM^\prime\sin(\phi_{q,L})\right)\right|=\left|\sin\left(\frac{\pi}{\lambda}dM^\prime\sin(\phi_{q,U})\right)\right|=\sin\left(\frac{\pi q}{M^\prime}\right)\alpha^\prime$. These angles follow the expressions
\begin{align}
\phi_{q,L}&=\arcsin\left(\tfrac{\lambda q}{M^\prime d}+\tfrac{\arcsin\left(\alpha^\prime\sin\left(\frac{\pi q}{M^\prime}\right)\right)}{\pi }\tfrac{\lambda}{M^\prime d}\right),\\
\phi_{q,U}&=\arcsin\left(\tfrac{\lambda(q+1)}{M^\prime d}-\tfrac{\arcsin\left(\alpha^\prime\sin\left(\frac{\pi q}{M^\prime}\right)\right)}{\pi }\tfrac{\lambda}{M^\prime d}\right)\nonumber.
\end{align}
Using this result, we rewrite the probability in \eqref{eq:prob} as 
\begin{align}
\tfrac{P\{i(\theta_j)<\alpha\}}{2}&\leq
\sum_{m=0}^{\frac{M^\prime-1}{2}}P\left\{\theta_j\in\left[l_{q,L},\phi_{q,L}\right]\right\}\nonumber\\
&+P\left\{\theta_j\in\left[\phi_{q,U},l_{q,U}\right]\right\}.
\label{eq:probFinal}
\end{align}
Note that $c(r_k,0)$ in \eqref{eq:equity} satisfies $\lim_{M\rightarrow\infty}c(r_k,0)=0$. Therefore, $M^\prime\approx 1$ in the asymptotic limit, and the right-hand side of \eqref{eq:probFinal} becomes $0$, regardless the chosen value of $\alpha$. It is therefore impossible to find a set of semi-orthogonal users. Furthermore, note that $\lim_{M\rightarrow\infty}c(r_k,\theta_k)=0$ also holds in the case $\theta_k\neq 0$. We note that $|\theta_k|>0$ leads to a smaller number of significant antennas $M^\prime$.

\bibliographystyle{IEEEtran}
\bibliography{references}

\end{document}